%% file: main.tex
\documentclass{article}
\usepackage{arxiv}
\usepackage{natbib}
\usepackage{graphicx,url}
\usepackage{xurl}
\usepackage{amsmath, listings}
\usepackage[utf8]{inputenc}
\usepackage[english]{babel}
\usepackage{minted}
\usepackage{hyperref}
\usepackage[per-mode = symbol]{siunitx}
\usepackage[acronym,nonumberlist,nogroupskip,noredefwarn]{glossaries}

\title{DualPI2 Active Queue Management in ns-3: Implementation and Validation}
\hypersetup{
  colorlinks,
  citecolor=black,
  filecolor=black,
  linkcolor=black,
  urlcolor=black,
  pdflang={en},
  pdftitle={DualPI2 Active Queue Management in ns-3: Implementation and Validation},
  pdfsubject={cs.NI},
  pdfauthor={Maria Eduarda Veras}
}

\author{%
    \\
    \textbf{Maria Eduarda Veras} \\
    Centro de Informática (CIn) \\
    Grupo de Pesquisa em Redes e Telecomunicações\\(GPRT)\\
    Universidade Federal de Pernambuco (UFPE) \\
    Recife, Brasil \\
    \texttt{eduarda.martins@gprt.ufpe.br}
    \and
    \\
    \textbf{Eduardo Freitas} \\
    Centro de Informática (CIn) \\
    Grupo de Pesquisa em Redes e Telecomunicações\\(GPRT)\\
    Universidade Federal de Pernambuco (UFPE) \\
    Recife, Brasil \\
    \texttt{eduardo.freitas@gprt.ufpe.br}
    \and
    \\
    \textbf{Assis T. de Oliveira Filho} \\
    Centro de Informática (CIn) \\
    Grupo de Pesquisa em Redes e Telecomunicações\\(GPRT)\\
    Universidade Federal de Pernambuco (UFPE) \\
    Recife, Brasil \\
    \texttt{assis.tiago@gprt.ufpe.br}
    \and
    \\
    \textbf{Djamel Sadok} \\
    Centro de Informática (CIn) \\
    Grupo de Pesquisa em Redes e Telecomunicações\\(GPRT)\\
    Universidade Federal de Pernambuco (UFPE) \\
    Recife, Brasil \\
    \texttt{jamel@gprt.ufpe.br}
    \and
    \\
    \textbf{Judith Kelner} \\
    Centro de Informática (CIn) \\
    Grupo de Pesquisa em Redes e Telecomunicações\\(GPRT)\\
    Universidade Federal de Pernambuco (UFPE) \\
    Recife, Brasil \\
    \texttt{jk@gprt.ufpe.br}
}
\renewcommand{\shorttitle}{}

\begin{document}
% \makeatletter
% \newcommand{\linebreakand}{%
%   \end{@IEEEauthorhalign}
%   \hfill\mbox{}\par
%   \mbox{}\hfill\begin{@IEEEauthorhalign}
% }
% \makeatother

\newacronym{l4s}{L4S}{Low Latency, Low Loss, and Scalable Throughput}
\newacronym{ran}{RAN}{Radio Access Network}
\newacronym{tcp}{TCP}{Transmission Control Protocol}
\newacronym{ietf}{IETF}{Internet Engineering Task Force}
\newacronym{rtt}{RTT}{Round-Trip Time}
\newacronym{3gpp}{3GPP}{3rd Generation Partnership Project}
\newacronym{rfc}{RFC}{Requests for Comments}
\newacronym{udp}{UDP}{User Datagram Protocol}
\newacronym{cca}{CCA}{Congestion Control Algorithm}
\newacronym{mss}{MSS}{Maximum Segment Size}
\newacronym{aqm}{AQM}{Active Queue Management}
\newacronym{ecn}{ECN}{Explicit Congestion Notification}
\newacronym{ect}{ECT}{ECN-Capable Transport}
\newacronym{cwr}{CWR}{Congestion Window Reduced}
\newacronym{ece}{ECE}{ECN-Echo}
\newacronym{ce}{CE}{Congestion Experienced}
\newacronym{red}{RED}{Random Early Detection}
\newacronym{pie}{PIE}{Proportional Integral controller Enhanced}
\newacronym{codel}{CoDel}{Controlled Delay Management}
\newacronym{fqcodel}{FQ-CoDel}{Flow Queue CoDel}
\newacronym{ai}{AI}{additive increase}
\newacronym{md}{MD}{multiplicative decrease}
\newacronym{aimd}{AIMD}{additive increase, multiplicative decrease}
\newacronym{dctcp}{DCTCP}{Data Center TCP}
\newacronym{ack}{ACK}{acknowledgment}
\newacronym{vpn}{VPN}{Virtual Private Network}
\newacronym{dualq}{DualQ}{Dual Queue Coupled AQM}
\newacronym{pi2}{PI2}{PI improved with a square}
\newacronym{wrr}{WRR}{Weighted Round Robin}
\newacronym{accecn}{AccECN}{More Accurate ECN Feedback}
\newacronym{ewma}{EWMA}{Exponentially Weighted Moving Average}
\newacronym{rack}{RACK-TLP}{Recent ACK and Tail Loss Probe}
\newacronym{dupack}{DupACK}{duplicate ACK}
\newacronym{srtt}{srtt}{smoothed RTT}
\newacronym{qos}{QoS}{Quality of Service}
\newacronym{qoe}{QoE}{Quality of Experience}
\newacronym{bdp}{BDP}{Bandwidth-Delay Product}
\newacronym{vm}{VM}{Virtual Machine}
\newacronym{nic}{NIC}{Network Interface Card}
\newacronym{quic}{QUIC}{Quick UDP Internet Connections}
\newacronym{ip}{IP}{Internet Protocol}
\newacronym{mtu}{MTU}{Maximum Transmission Unit}

\maketitle

\begin{abstract}
The demand for ultra-low latency applications necessitates advanced network architectures like the \gls{l4s} standard. A core component of \gls{l4s} is the DualPI2 \gls{aqm}, which ensures the safe coexistence of scalable and classic traffic. Despite \gls{l4s}'s growing adoption, the ns-3 network simulator lacks a high-fidelity, up-to-date DualPI2 model. This paper presents a comprehensive implementation of the DualPI2 \gls{aqm} in ns-3, while also mirroring the official Linux Kernel architecture. Our model incorporates representative mechanisms previously absent in simulation, such as credit-based \gls{wrr} scheduling, step-marking, and overload protection. To guarantee simulation accuracy, we validate our implementation against a physical Linux testbed across 25 diverse Bandwidth-Delay Product (BDP) scenarios. Results demonstrate that our ns-3 model replicates real-world behavior, ensuring strict throughput fairness and queue delay isolation. Ultimately, this validated model equips the research community with a robust tool to evaluate and advance \gls{l4s} performance across diverse network topologies, ranging from data centers to wireless home and office environments.
\end{abstract}

% \begin{IEEEkeywords}
% ns-3, Active Queue Management, DualPI2, L4S, Network Simulation
% \end{IEEEkeywords}

\glsresetall

\section*{Publishing}
This is an earlier version of the paper submitted and accepted to SoftCOM 2026 conference. Once the paper is accepted, we will update our arXiv repository.

\section{Introduction}
\label{sec:introduction}
\input{introduction}

\section{Theoretical Background}
\label{sec:background}
\input{background}

\section{Related Work}
\label{sec:related-work}
\input{rel-work}

\section{Implementation}
\label{sec:implementation}
\input{implementation}

\section{Evaluation}
\label{sec:evaluation}
\input{evaluation}

\section{Results}
\label{sec:results}
\input{results}

\section{Conclusion and Future Work}

This article presented an implementation of DualPI2 \gls{aqm} for the ns-3 network simulator. By adhering to the architectural logic of the official Linux DualPI2 \texttt{qdisc}, incorporating mechanisms such as credit-based \gls{wrr} scheduling, fallback overload protection, and step marking, we addressed a significant gap in the current simulation environment.

The accuracy of the proposed model was validated in a Linux physical test environment. In 25 diverse \gls{bdp} scenarios, the ns-3 implementation reproduced real-world behavior. The empirical results demonstrate the model's ability to accurately isolate ultra-low latency scalable streams while maintaining throughput fairness with classic traffic.

Ultimately, this validated model provides the networking research community with a system-level tool for evaluating the \gls{l4s} architecture across diverse topologies and platforms, such as Wi-Fi, 4G/5G, and data center environments, which are often impractical to deploy at scale in physical test environments.

Future work will focus on the development and integration of a fully compliant TCP Prague congestion control module for ns-3. This addition will complete the \gls{l4s} simulation suite, enabling comprehensive, end-to-end evaluations of scalable congestion controls and their adherence to the Prague Requirements under highly dynamic network conditions.

\section*{Software Availability}
To foster reproducible networking research, the complete ns-3 DualPI2 source code, along with the scripts used to generate the 25 evaluation scenarios, are publicly available at: \url{https://github.com/GPRT/l4s-for-ns3}.

\section*{Acknowledgments}
This work was supported by Ericsson Telecomunicações Ltda., and by the São Paulo Research Foundation (FAPESP), grant 2021/00199-8, CPE SMARTNESS.

\bibliographystyle{unsrt}
\bibliography{references}

\end{document}

%% file: introduction.tex
% -*- TeX-+master:"./main.tex"; auto-fill-function:nil; -*-

Over the last decade, the requirements for network services have evolved beyond demanding high bandwidth, to also rely on ultra-low latency. Interactive services, such as cloud gaming, augmented reality, videoconferencing, and autonomous vehicle control, depend on minimal response times to guarantee \gls{qoe} \cite{rfc9330}. The rapid growth of these applications requires the underlying network technologies capable of supporting strict timing constraints \cite{Cisco2020cisco-annual-internet-report-2018-2023}.

However, transmissions can suffer from extreme high latency scenarios on the Internet, even when overcoming physical propagation delay. One of the main causes of these scenarios is \textit{queue delay}, which represents how long a packet remains in a queue buffer on network nodes such as routers, prior to be forwarded. The longer a packet has to wait, the higher the queue delay will be, thus increasing the overall end-to-end packet latency.

To mitigate this excessive queue delay, Internet engineers created algorithms that control the number of packets occupying the queue. These algorithms, called \gls{aqm}, proactively control buffer occupancy by dropping packets before queue saturation \cite{rfc7567}. As a consequence, this packet drop acts as a congestion signal to the sender, usually a TCP sender, forcing it to reduce its sending rate. Furthermore, \glspl{aqm} were able to prevent packet loss and explicitly notify senders of congestion by using the \gls{ecn} mechanism. With \gls{ecn}, \glspl{aqm} can signal congestion directly in the IP header, allowing senders to adjust their transmission rates without relying on packet drops \cite{rfc3168}.

However, these mechanisms cannot easily provide ultra-low latency traffic for traditional flows. This is not due to the algorithms used in the \glspl{aqm}, but instead has to do with the way transmission protocols control network congestion and bandwidth probing. These protocols, such as TCP or QUIC, deploy a \gls{cca}, a mechanism responsible for continuously discover the available bandwidth it is supposed to use. The \gls{cca}, however, is also responsible for generating an oscillating pattern of bandwidth probing that inherently induces queue delay at the bottleneck \gls{aqm}.

To address this limitation, the \gls{l4s} architecture introduces two core components to enable safe, low-latency traffic coexistence. The first is a \textit{scalable} \gls{cca} designed to probe bandwidth without inducing queue delay. The second is a new \gls{aqm} algorithm, the \gls{dualq}, which isolates latency-sensitive flows from classic, capacity-seeking traffic while maintaining throughput fairness.

\gls{l4s} adoption continues to expand across diverse access technologies, including Wi-Fi, DOCSIS, and 5G networks \cite{tmobile,cablelabs}. Despite this momentum, a formal \gls{l4s} implementation remains absent in the ns-3 simulator. ns-3 is one of the main network simulators available in the literature, being a state-of-the-art experimentation tool for diverse networking environments. Integrating \gls{l4s} protocols into ns-3 allows for conducting reproducible experiments across diverse topologies, from data centers to wireless, and for evaluating the coexistence of emerging protocols.

Given this context, this paper presents a high-fidelity implementation of the \gls{dualq} in the ns-3 simulator. Unlike previous models, our work incorporates the Linux implementation of the DualPI2 qdisc scheduler, ensuring that simulation mirror the behavior of real-world network stacks. To validate the accuracy of our model, we validate the simulation results against a physical testbed running the Linux implementation.

The remainder of this paper is organized as follows: Section II provides the theoretical background on L4S and DualPI2. Section III reviews related work. Section IV details the ns-3 implementation, followed by the validation methodology in Section V. Section VI discusses the results. Finally, Section VII concludes the paper.

%% file: background.tex
% -*- auto-fill-function:nil; -*-

As we briefly mentioned, \gls{l4s} has two major components, the scalable \gls{cca} and the \gls{dualq}. Although it is a simplification -- these two components are complex and provide a set of rigorous requirements to enable safe, low-latency traffic coexistence -- \gls{l4s} indeed depends on these two components, and one cannot be deployed without the other.

A scalable congestion control is a new category of \glspl{cca} that does not induce queue delay on the network to probe bandwidth. It achieves this by receiving constant \gls{ecn} marks from the network and scaling its window proportionally to the amount of these \gls{ecn}-marked packets. This allows the congestion control to adapt to the network capacity without having packets in a standing queue, given that the queue will mark the packets as soon as congestion starts. In addition, these \glspl{cca} will also be served in a ``shallow-buffer'' queue, meaning that since it does not induce queue delay, it can safely make use of smaller queues to further guarantee its low queue delay.

The first scalable congestion control to be widely adopted in real-world scenarios was \gls{dctcp}. As the name implies, it is designed for data centers, since its behavior can be aggressive for other unscalable \glspl{cca} -- often referred as classic \glspl{cca}. Since data centers are a controlled environment, administrators can configure all hosts and network nodes to properly support \gls{dctcp}, preventing such aggressiveness issues. In fact, \gls{l4s} originated as an initiative to enable \gls{dctcp} to exist across the broader Internet \cite{7063495} and the component that could enable this was an appropriate \gls{aqm}. The architecture subsequently evolved beyond \gls{dctcp}, with the standardization of the ``Prague Requirements'' \cite{rfc9330}, a set of safety criteria that any scalable \gls{cca} must satisfy to operate safely on the Internet without starving classic flows. Indeed, \gls{dctcp} is such an example of scalable \gls{cca} that does not adhere to the Prague Requirements, and because of this, it is not considered an \gls{l4s} \gls{cca}. Conversely, TCP Prague was designed to meet these requirements, and it is the official \gls{l4s} protocol from the \gls{l4s} researchers \cite{briscoe-iccrg-prague-congestion-control-04}.

However, the careful reader may have realized that scalable congestion control relies heavily on the queue management from the network. We mentioned that it expects frequent \gls{ecn} marks from the network and also a shallow-buffer \gls{aqm} to keep its packets the shortest time possible inside the queue. Therefore, a scalable \gls{cca} cannot operate in a standard \gls{aqm} with large buffers. In contrast, classic flows still need these standard \glspl{aqm}, since they need larger buffers to accommodate their delay-inducing traffic. In essence, scalable and classic \glspl{cca} are fundamentally incompatible and cannot coexist on the Internet without causing starvation issues.

To resolve this incompatibility, \gls{l4s} standardized the \gls{dualq} \gls{aqm}. Related studies demonstrate that isolating the two \gls{cca} categories into separate queues is sufficient for coexistence, provided that the congestion signaling (marking and/or dropping) is mathematically coupled between them.

The \gls{dualq} architecture operates through three primary stages. The first component is the \textit{classifier}, which identifies the traffic as being \gls{l4s} or classic traffic. This identification uses the \gls{ecn} field on the IP header as standardized in RFC 9332 \cite{rfc9332}. It establishes that only \gls{l4s} traffic uses the \gls{ect}(1) bits, while classic flows can use either \gls{ect}(0) or Not-ECT.

After identification, the classifier enqueues the packets into their respective buffers, where the second stage, \textit{coupled congestion signaling}, occurs. The innovation of \gls{dualq} derives  the marking and dropping probabilities for both queues from a single, shared base probability \cite{albisser2019dualpi2}. This base probability is mathematically scaled up for the \gls{l4s} queue to deliver the frequent \gls{ecn} marks expected by scalable \glspl{cca}, and scaled down for the classic queue, which requires fewer signaling. However, both probabilities are mathematically coupled, resulting in a bandwidth probing for both traffic types that is fair and prevents the starvation of any of the two traffic classes. 

\begin{figure}[htb]
    \centerline{\includegraphics[width=0.7\textwidth]{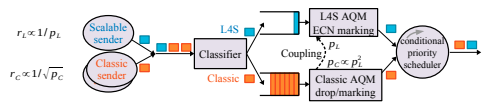}}
    \caption{Structure of the DualPI2 AQM \cite{implementing-prague}.}
    \label{fig:aqm}
    % TODO: If time is available after reviews, we should make our own version of this image
\end{figure}

The final component of the \gls{dualq} is the conditional priority scheduler, which will dequeue \gls{l4s} packets with higher frequency than classic, while ensuring that classic traffic does not starve. Figure~\ref{fig:aqm} illustrates the \gls{dualq} architecture.

While \gls{dualq} is an architectural concept applicable to any \gls{aqm}, the DualPI2 algorithm \cite{albisser2019dualpi2} is its primary and most developed implementation. DualPI2 applies the coupling mechanisms from the PI2 \gls{aqm}, a modified version of the state-of-the-art PIE \gls{aqm}. In DualPI2, a base probability $p'$ is used as the probability to drop or mark with \gls{ecn} a given packet. $p'$ is derived from the base PIE algorithm, which uses a proportional-integral equation to control the queue delay, as denoted in Equation~\ref{eq:pie-alg}. If the current queue delay exceeds the target and/or the previous queue delay measurement, $p'$ increases -- marking/dropping more packets -- to signal congestion to the senders and reduce sending rates.

To achieve the required coupled signaling, this base probability is scaled according to the queue. For the classic queue, the classic probability ($p_C$) is defined as the square of the base probability, effectively reducing the signaling frequency. For the \gls{l4s} queue, the L4S probability ($p_L$) is multiplied by a ``coupling factor'' $k$ -- defined as 2 --, which increases the signaling rate. A rationale behind why these values are used in available in \cite{deschepper2022dualqueuecoupledaqm}.

%\begin{equation}
%  p' = p_{prev} + \alpha(del_{curr} - del_{target}) + \beta(del_{curr} - del_{prev})
%  \label{eq:pie-alg}
%\end{equation}
%\[
%  p_C = (p')^2
%\]
%\[
%  p_L = p' \times k
%\]

\[
  p' = p_{prev} + \alpha(delay_{cur} - delay_{target}) + \beta(delay_{cur} - delay_{prev})
\]
\begin{equation}
  p_C = (p')^2 \text{  and  } p_L = p' \times k
  \label{eq:pie-alg}
\end{equation}

%% file: rel-work.tex
% -*- TeX-+master:"./main.tex"; auto-fill-function:nil; -*-

Despite the architecture of the DualPI2 \gls{aqm} being introduced and extensively evaluated within the Linux kernel by De Schepper et al. \cite{deschepper2022dualqueuecoupledaqm}, the translation of these concepts into high-fidelity simulation models remains limited. Most existing efforts within the ns-3 ecosystem are confined to unofficial codebases and community drafts rather than peer-reviewed publications. In our previous work \cite{11197397}, where we seek to provide \gls{l4s} traffic prioritization on 5G networks, we observed this gap on the literature.

The only formal published implementation for the simulator is the single-queue PI2 algorithm by Tahiliani, in 2017 \cite{pi2_tahiliani}. While this work presented an implementation of the base proportional-integral enhanced with a square controller logic for ns-3, it strictly operates as a single-queue mechanism. Since it lacks the DualQ architecture, it is unable to maintain low queue delay for scalable flows, the fundamental requirement for \gls{l4s}. Furthermore, this contribution was never merged into the official mainline, and has since received no updates.

Community efforts to extend this PI2 model into a dual-queue architecture began with Shravya KS's 2017 Google Summer of Code project \cite{Shravya2017GSoC}, which was later refined by Tom Henderson in an unofficial repository \cite{HendersonDualQ} in 2019. Although this implementation correctly translates the DualQ architecture pseudo-code in RFC 9332 \cite{rfc9332}, it lacks the structural optimizations that DualPI2 presents, especially when analyzing the Linux kernel version. For instance, it uses a simpler Time-Shifted FIFO (TS-FIFO) scheduler instead of the standard credit-based \gls{wrr}. It also omits essential control mechanisms, such as the fallback overload protection logic and step-driven \gls{ecn} marking (\texttt{StepAqm}). Because it serves primarily as an experimental baseline, this codebase was also never merged into the official ns-3 releases.

An alternative conceptual approach for handling \gls{l4s} traffic is leveraging flow-queuing, specifically configuring FQ-Codel with a shallow-buffer threshold for \gls{ect}(1) packets \cite{fqcodel-l4s}. While this provides partial latency benefits by segregating scalable flows, it inherently abandons the mathematically coupled probability signaling that defines the \gls{dualq} paradigm. Without strict coupling, this approach yields sub-optimal fairness and coexistence dynamics compared to a dedicated DualPI2 \gls{aqm} \cite{deschepper2022dualqueuecoupledaqm}.

The gap in the current literature is therefore clear: existing simulation efforts either lack the dual-queue architecture entirely, substitute coupled signaling with flow isolation, or rely on theoretical simplifications that fail under complex traffic loads. Our work directly addresses this by shifting the design paradigm from a specification-based model to an implementation-based one. By porting the scheduling logic, edge-case protections, and structural optimizations of the Linux DualPI2 \texttt{qdisc} into ns-3, we propose a mathematically rigorous and system-accurate tool. This ensures that researchers can evaluate \gls{l4s} performance exactly as it manifests in real-world deployments.

%% file: implementation.tex
% -*- auto-fill-function:nil; -*-

The implementation of the DualPI2 \gls{aqm} in ns-3 was carried out using version 3.47 of the simulator, focusing specifically on the \texttt{traffic-control} module. While the related work in \cite{HendersonDualQ} provided an initial foundation for basic dual-queue mechanisms, its codebase required additional updates to ensure compatibility with modern releases. Furthermore, to follow the official Linux kernel implementation present in \cite{l4s_linux_kernel}, multiple modifications were necessary, including enqueueing and dequeueing procedures, drop management, overload handling, and a new conditional scheduler, while preserving standard \texttt{QueueDisc} functions like \texttt{CheckConfig} and \texttt{IsL4S}. This section details these core modifications.

The first component in the \gls{aqm} is the packet arrival pipeline within the \texttt{DoEnqueue} method. Incoming packets are classified via the IP header's \gls{ecn} field: ECT(1) traffic is directed to the \gls{l4s} queue, while ECT(0) or Not-ECT goes to the Classic queue. Prior to buffering, the packet undergoes two evaluations. First, following the Linux \textit{Step \gls{aqm}}, the \gls{l4s} queue depth is checked against a packet threshold. If exceeded, we attach an ns-3 \texttt{Tag} to the packet, emulating the Linux \texttt{skb} flag, to authorize marking during dequeue. Second, if the \texttt{drop\_early} variable is active, \texttt{MustDrop} probabilistically discards the packet before it consumes memory. Packets passing these checks are enqueued.

The \texttt{MustDrop} function manages probabilistic discarding and mitigates severe congestion. To protect small flows, it avoids drop if the combined queue size is less than 2 \glspl{mtu}. It then calculates the \gls{l4s} and Classic probabilities from the base probability, as explained in Section~\ref{sec:background}. An overload state triggers when $p_L$ exceeds 100\%. Because \gls{ecn} signaling cannot scale beyond this limit, the \gls{aqm} enforces strict drops to relieve buffer pressure. During overload, \gls{l4s} packets, normally marked using $p_L$, are evaluated against $p_C$ and dropped if the \texttt{drop\_overload} flag is active. Meanwhile, classic packets attempt to receive an \gls{ecn} mark based on $p_C$, but are immediately dropped if the system is in overload. If \texttt{drop\_overload} is not active, marking will continue following the $p_L$ and $p_C$ probability, trading drop rate for queue delay \cite{deschepper2022dualqueuecoupledaqm}.

Congestion probabilities are updated dynamically by the \texttt{DualPi2Update} function. Executing every 16 milliseconds, following the default Linux implementation, this routine computes the shared base probability, $p'$. We consider the current queue delay as the maximum head delay between the \gls{l4s} and classic queue. This delay drives the PI controller to calculate the raw $p'$ value, following Equation~\ref{eq:pie-alg}. Finally, if the \texttt{drop\_overload} mechanism is disabled, the function clamps $p'$ to a maximum of $1.0 / k$. This bounding ensures the derived \gls{l4s} probability never exceeds \qty{100}{\percent}, maintaining congestion window fairness during severe queue overflows, at the expense of increasing queue delay.

The final component relies on the credit-based \gls{wrr} scheduler and the \texttt{DoDequeue} method. To meet \gls{l4s} latency targets without starving Classic traffic, the scheduler enforces a 90:10 weight ratio. While this specific proportion matches the Linux default, the exact scheduler weight is immaterial for capacity sharing, which is instead dictated by the \gls{l4s} senders' response to the coupled congestion signals \cite{deschepper2022dualqueuecoupledaqm}. Packet selection is conditional: an \gls{l4s} packet is selected if its buffer has items and either the classic queue is empty or \gls{l4s} credits remain. Otherwise, a classic packet is selected, and \gls{l4s} credits are replenished. If both queues are empty, the credit counter resets.

Once selected, the packet enters \texttt{DoDequeue}. At this stage, the \texttt{StepAqm} function evaluates \gls{l4s} packets for latency violations. If a packet carries the \texttt{Tag} (applied during enqueue) and its individual sojourn time exceeds a secondary delay threshold, it receives an immediate \gls{ecn} mark. Following this step logic, \texttt{MustDrop} executes any final probabilistic decisions. The scheduler’s credit variable is updated only \textit{after} these functions return, ensuring that \gls{l4s} credits are never wasted on packets that are ultimately dropped.

%% file: evaluation.tex
% -*- auto-fill-function:nil; -*-

We evaluated our DualPI2 model using a standard dumbbell topology configured for downlink traffic. The edge access links were over-provisioned (\qty{1}{\giga\bit\per\second}, \qty{0}{\milli\second} delay) to isolate congestion strictly at the central bottleneck. To assess the \gls{aqm} performance across a wide range of Bandwidth-Delay Products (BDP), we evaluated 25 parameter combinations, varying the bottleneck capacity (4, 12, 40, 120, and \qty{200}{\mega\bit\per\second}) and the base link \gls{rtt} (5, 10, 20, 50, and \qty{100}{\milli\second}). The DualPI2 discipline was installed on the server-side router's egress interface, managing all traffic entering the bottleneck with default parameters.

We deployed the \texttt{BulkSendApplication} to generate competing long-lived flows for 60 seconds. Each of the 25 scenarios was replicated 30 independent times, varying only the simulations's pseudo-random number generator (RNG) seed. To evaluate the core \gls{l4s} requirements, we extracted the following metrics: throughput and congestion window (\texttt{cwnd}) to assess fair bandwidth utilization; TCP retransmissions to measure packet loss; and sojourn time (queue delay) to verify ultra-low latency. In the ns-3 environment, these metrics were collected utilizing native \texttt{TcpSocketBase} trace sources to monitor the cwnd, and TCP retransmissions. At the bottleneck router, native queue traces logged packet marks and drops, and a custom trace source was added to measure the sojourn time in each sub-queue.

To establish a ground-truth baseline, we replicated all simulated scenarios in a physical Linux testbed equipped with a Gigabit Ethernet \glspl{nic}. The setup comprises two dedicated hypervisors, each hosting two virtual machines (Ubuntu 22.04, 1\,vCPU Intel Xeon E-2434, \qty{4}{\gibi\byte} RAM) acting as clients and servers. These instances communicate through two physical routers (Debian 13, Intel Core i5 3.33\,GHz 4-core, \qty{8}{{\gibi\byte}} RAM) functioning as the bottleneck. The routers executed the custom Linux kernel version 6.6.114-a76b708b2-l4steam-116, provided by the official \gls{l4s} repository \cite{l4s_linux_kernel}, to run the native DualPI2 \texttt{qdisc}. Testbed metrics were captured using standard tc qdisc iterations and the output of iperf software.

We have mentioned that \gls{dctcp} is not an L4S congestion control per se, even though it is scalable. We chose DCTCP as our scalable congestion control since TCP Prague is not yet officially available in ns-3. Part of our research aims to implement such a protocol, but it is not ready yet for experimentation. Furthermore, this approach directly mirrors the baseline validation methodology originally employed by the \gls{l4s} researchers during the conception of DualPI2 \cite{albisser2019dualpi2}. This is considered a valid approach, especially because the DualQ itself is designed to support scalable and classic CCAs regardless of its L4S compliance. Stated differently, from the AQM perspective, it can provide queue management fairness even though a real-world L4S scenario requires L4S-compliant CCAs to ensure Internet coexistence -- regarding non-AQM-related issues like RTT dependence \cite{albisser2019dualpi2}.

%% file: results.tex
% -*- TeX-+master:"./main.tex"; auto-fill-function:nil; -*-

\begin{figure*}[htb]
    \centerline{\includegraphics[width=\textwidth]
      {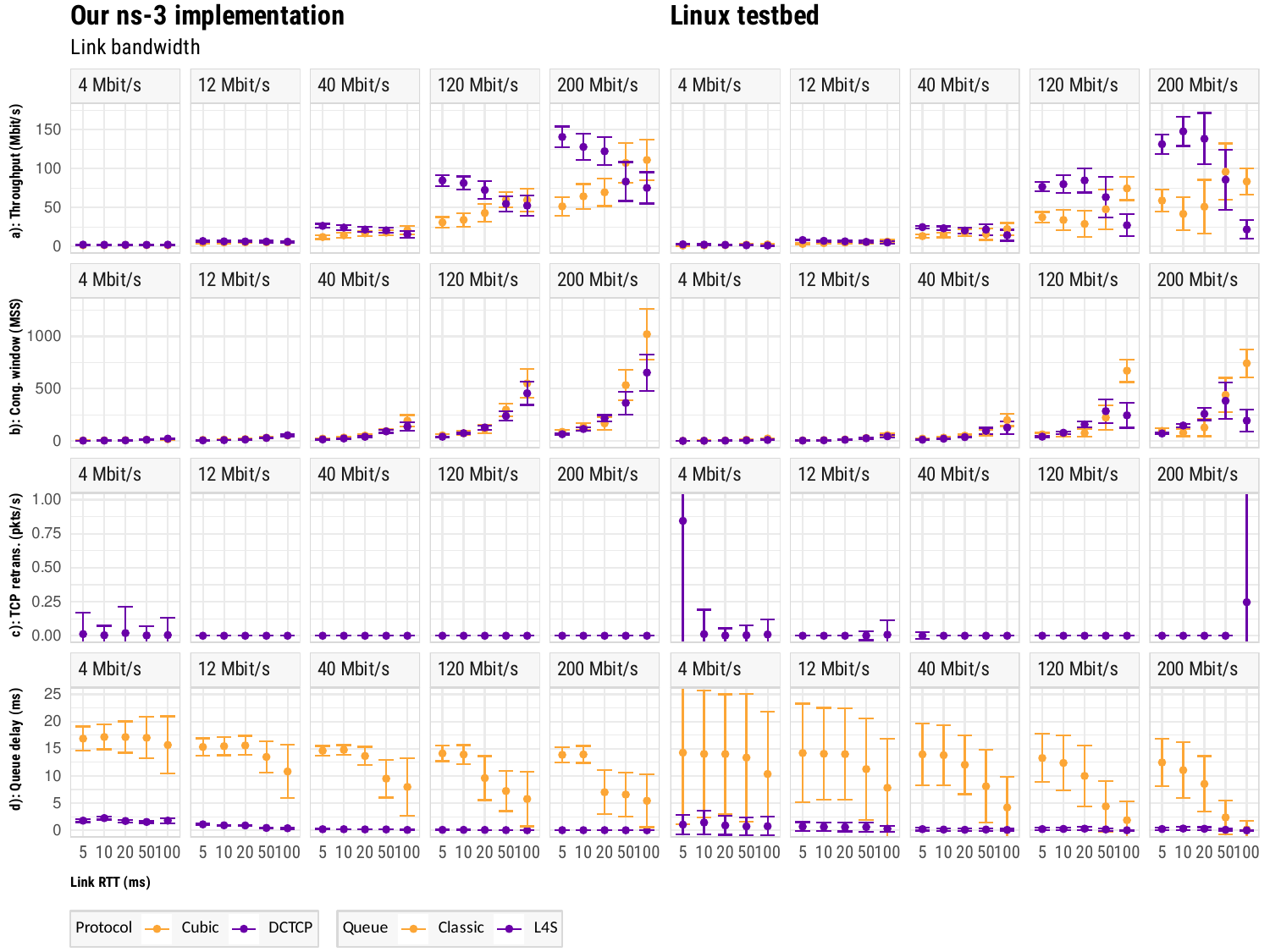}}
    \caption{Results for our ns-3 DualPI2 implementation against our testbed. All graphs show the mean with standard deviation for all 30 replications.}
    \label{fig:results}
\end{figure*}

Figure~\ref{fig:results}(a) shows the throughput, which is a direct way of analyzing bandwidth utilization fairness. We can see that DualPI2 enables fair utilization in the lower bandwidth scenarios of 4, 12, and \qty{40}{\mega\bit\per\second} with near-identical results from our implementation and the Linux testbed. DCTCP and Cubic share the same throughput with small variation. In the higher capacity scenarios of 120 and \qty{200}{\mega\bit\per\second}, we can see a slight difference on how the two protocols behave. In the first three link \gls{rtt} values, DCTCP uses more bandwidth than Cubic, as expected, since DCTCP tends to be more aggressive as DualPI2 prioritizes it. This behavior happens also in the Linux testbed.

However, we observe a different pattern at higher bandwidth and \gls{rtt} values, especially \qty{100}{\milli\second}. While \gls{dctcp} and Cubic throughput sit at \qty{25}{\mega\bit\per\second} and \qty{75}{\mega\bit\per\second} respectively in Linux, the ns-3 simulation shows both flows sharing the \qty{120}{\mega\bit\per\second} link evenly at around \qty{50}{\mega\bit\per\second} each. This discrepancy occurs because the Linux \gls{dctcp} flow remains constantly in the additive increase phase, struggling to reach maximum bandwidth utilization. This stagnation is driven by a rounding artifact in the Linux kernel known as \textit{cwnd latching}. As noted by the developers of the ns-3 \gls{dctcp} model, at larger \gls{rtt} values, the Linux window reduction only affects the integer part of the calculation, causing the window to become nearly stationary \cite{10.1145/3389400.3389405}. Since ns-3 abstracts these integer arithmetic limitations, its \gls{dctcp} model is more aggressive, bypassing this issue and achieving maximum throughput much faster, independently of our \gls{aqm} implementation.

\begin{figure*}[htb]
    \centerline{\includegraphics[width=\textwidth]
      {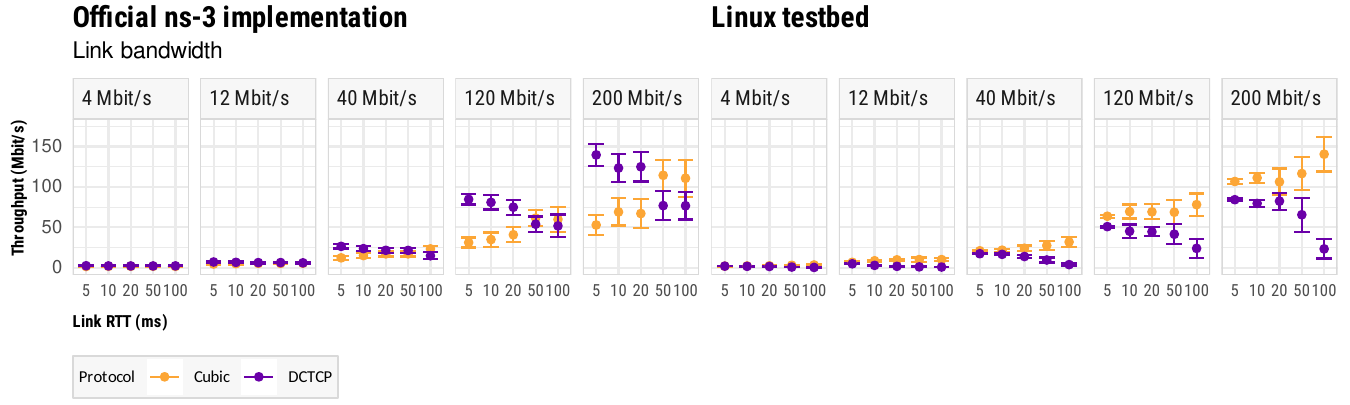}}
    \caption{Simple proof of concept of DCTCP and Cubic over the official FQ-CoDel ns-3 implementation against our testbed with the same queue. All graphs show the mean with standard deviation for all 12 executions.}
    \label{fig:results-fqcodel}
\end{figure*}

To validate our assumption, we made a simple experiment, replicating the same scenario with the FQ-CoDel \gls{aqm}, since it has also an official ns-3 implementation code irrespective of our modifications. To further improve our results, we activated \texttt{ce\_threshold}. Figure~\ref{fig:results-fqcodel} illustrates this replication. The results show that indeed, the Linux \gls{dctcp} implementation has a less aggressive behavior, staying well below the fair throughput when competing with Cubic on these higher link bandwidth and \gls{rtt} scenarios. On ns-3, not only it achieves higher throughput, but it also keeps a fair share on the \qty{100}{\milli\second} scenarios. With these experiments, we understand that the observed differences in the results of our DualPI2 \gls{aqm} are acceptable, especially as it does not indicate flow starvation.

Furthermore, TCP retransmission is also successfully replicated in our scenario (Figure~\ref{fig:results}(c)). The low bandwidth scenario has inevitably higher rates. Our implementation presents close values with the Linux testbed. In the remaining bottleneck scenarios, ns-3 is able to perform no TCP retransmissions, while Linux still has some outlier drops in the case of 12 Mbps or 200 Mbps. These retransmissions do not invalidate the low loss aspect of L4S, especially because they are isolated, which explains why the mean is less than 1 packet. We only show the retransmission for DCTCP, since Cubic is expected to have constant retransmission due to receiving drops as congestion signals.

Finally, we examine the main metric, queue delay in Figure~\ref{fig:results}(d). We show the classic queue delay as a reference to further ensure that fair bandwidth utilization is achieved even though queue delay is drastically different. However, our main concern is the L4S queue. Results indicate that the mean queue delay for L4S is always below 2 ms with low variation. In scenarios of 4 and 12 Mbps, it is expected to have higher delays since such small BDPs will increase the serialization time at the queue, so we increase the L4S step AQM threshold to accommodate such longer times. This isolation of latency-sensitive traffic, achieved without degrading the throughput of classic flows, validates our implementation.

